\documentclass[sigconf]{acmart}
\AtBeginDocument{%
  \providecommand\BibTeX{{%
    \normalfont B\kern-0.5em{\scshape i\kern-0.25em b}\kern-0.8em\TeX}}}

\setcopyright{acmcopyright}
\copyrightyear{2023}
\acmYear{2023}
\acmDOI{XXXXXXX.XXXXXXX}

\acmConference[CoNEXT-SW '23]{ACM on Networking (PACMNET)}{December 8,
  2023}{Paris, France}
\acmISBN{978-1-4503-XXXX-X/23/12}

\begin{document}

\title{GNN-based Anomaly Detection for Encoded Network Traffic}

\author{Anasuya Chattopadhyay}
\email{anasuya.chattopadhyay@dfki.de}
\orcid{0009-0005-4296-3440}
\affiliation{%
  \institution{Intelligent Networks Research Group, German Research Center for Artificial Intelligence}
  \city{Kaiserslautern}
  \country{Germany}}

\author{Daniel Reti}
\email{daniel.reti@dfki.de}
\orcid{0000-0001-8071-6188}
\affiliation{%
  \institution{Intelligent Networks Research Group, German Research Center for Artificial Intelligence}
  \city{Kaiserslautern}
  \country{Germany}}

\author{Hans D. Schotten}
\email{schotten@eit.uni-kl.de}
\affiliation{%
  \institution{Intelligent Networks Research Group, German Research Center for Artificial Intelligence}
  \city{Kaiserslautern}
  \country{Germany}}

\renewcommand{\shortauthors}{Chattopadhyay, et al.}

\begin{abstract}
The early research report explores the possibility of using Graph Neural Networks (GNNs) for anomaly detection in internet traffic data enriched with information. While recent studies have made significant progress in using GNNs for anomaly detection in finance, multivariate time-series, and biochemistry domains, there is limited research in the context of network flow data. In this report, we explore the idea that leverages information-enriched features extracted from network flow packet data to improve the performance of GNN in anomaly detection. The idea is to utilize feature encoding (binary, numerical, and string) to capture the relationships between the network components, allowing the GNN to learn latent relationships and better identify anomalies.

\end{abstract}





\maketitle

\section{Introduction}

Anomaly detection plays a critical role in cybersecurity, where identifying unusual or malicious activities on the internet is of paramount importance. Traditional anomaly detection methods often rely on predefined rules and heuristics, making them less adaptable to ever-evolving threats. Graph Neural Networks (GNNs) offer a promising avenue for improving the detection of such threats by capturing complex relationships and patterns in network data. A typical GNN architecture consists of multiple layers, each updating node representations based on their neighbors' information. The final node embeddings are then used for anomaly detection. Post-training, GNNs can be deployed for real-time anomaly detection by analyzing the incoming traffic data and flagging instances that deviate from learned patterns as potential anomalies.

Recent research in anomaly detection has predominantly focused on domains like image \cite{han-gnn-image}, finance \cite{jullum-gnn-finance}, Internet of Things (IoT) \cite{wu-gnn-iot, DeMedeiros-gnn-iot}, multivariate time-series \cite{han-gnn-multivariate, Deng_Hooi_gnn-multivariate}, and biochemistry \cite{li-gnn-bio, qiu-gnn-bio}. These applications have shown significant success with GNNs. Similarly in the area of Network Security, the application of GNNs is used for Intrusion detection (ID) \cite{caville-gnn-ad} and attacks on the network \cite{casas-gnn-ad}. However, the detection of anomalies in the packet level for network traffic remains unexplored. In this early research report, we propose an approach by incorporating information-enriched features derived from the packet level into a GNN-based anomaly detection framework.

\subsection{Problem statement}

Internet traffic contains a vast amount of information, and distinguishing between normal and malicious activities is challenging. Traditional rule-based approaches often fail to adapt to evolving cyber threats, leading to security vulnerabilities. It is crucial to develop a robust and adaptive anomaly detection system that can identify unusual network behaviors and potential threats in real time.

Effective anomaly detection is essential to identify and mitigate cyber threats, safeguarding sensitive data and critical infrastructure. Some known challenges in anomaly detection could be presence of bias in training data. Biases can emerge in anomaly detection, particularly when training data is not representative of the entire spectrum of cyber threats. Common biases include over-representation of certain attack types and under-representation of others, leading to model performance limitations. GNNs may have the potential to mitigate these bias by learning from data and adapting to new and previously unseen attack patterns. Another challenge is the low coverage of various attacks in the training data. GNNs may perform sub-optimally when faced with previously unseen attack patterns. GNNs might identify novel and sophisticated attack patterns that may not be captured by traditional rule-based systems. Often the training data lacks information on changing network environments and emerging threats and graph-based learning might have the potential to fill the gap.

\section{Proposal}

The proposed idea involves the use of Graph Neural Networks (GNNs) to model the complex relationships within internet traffic data. Network flow data consists of records that describe communication sessions between devices or hosts in a computer network. Each flow record typically includes information such as source and destination IP addresses, port numbers, protocol details, packet count, and byte count. To enhance anomaly detection in internet traffic data, we propose feature encoding, which involves extracting meaningful attributes from network flow data. 
Feature encoding can capture complex relationships between entities. For example, encoding the source and destination IP addresses, protocol details, and other attributes as embeddings might help the GNN learn the relationships between these components within the network. The network flow data is represented as a graph, where nodes correspond to devices or hosts, and edges represent communication sessions. The GNN architecture is trained on labeled data to distinguish between normal and anomalous network behaviors.

\section{Hypotheses}

The expected effects of the proposed solution are:
\begin{itemize}
    \item \textbf{Improved Anomaly Detection}: GNNs, with their ability to capture complex relationships, are expected to outperform traditional methods in identifying network anomalies.
    \item \textbf{Enhanced Adaptability}: The GNN-based approach is expected to adapt better to evolving cyber threats due to its ability to learn from historical and real-time data.
\end{itemize}

Plausible alternatives to the proposed solution include:
\begin{itemize}
    \item Traditional Rule-Based Approaches: These methods rely on predefined rules and heuristics. While they are interpretable, they may struggle to handle complex, evolving threats.
    \item Other Machine Learning Models: Alternative machine learning models, such as decision trees or random forests, may be considered. However, these models may not capture intricate network relationships as effectively as GNNs.
\end{itemize}

\section{Experiments}

To test the hypotheses and achieve desired results, we will start by collecting a diverse dataset of network flow data with labeled anomalies and normal behaviors from public sources (like OpenML), for example, Burst Header Packet Dataset, KDDCup99, BOT-IoT, KDD\textunderscore Internet\textunderscore Usage, Internet-firewall, and PhishingWebsites. Subsequently, we will implement feature extraction and encoding techniques tailored to the data type, ensuring alignment with the specific dataset. Additionally, we will carefully choose hyperparameters like activation function, network architecture, and initialization methods to prevent issues like hypersphere collapse \cite{zhao-hypersphere}, which can hinder the model's ability to capture intricate data patterns. After training our model, we will compare its performance with existing benchmarks for network flow data and evaluate its effectiveness across various datasets. To assess the GNN's anomaly detection performance, we will utilize standard metrics such as precision, recall, and F1-score analysis. These experiments will validate our hypotheses by showcasing the GNN's capability to differentiate between normal and anomalous network behaviors, with conclusions grounded in empirical evidence derived from labeled data.

\section{Results and Conclusion}
In conclusion, this technical report has presented an approach in which we can use Graph Neural Networks (GNNs) for anomaly detection in internet traffic data, focusing on the cybersecurity domain. By enriching network flow data with relevant features and representing it as a graph, GNNs have the potential to learn complex relationships and patterns, which can result in accurate anomaly detection. The project's success will be determined based on qualitative analysis of the graph embeddings and the inferences that GNN make, ultimately contributing to improved cybersecurity defenses against network-based threats.


\bibliographystyle{ACM-Reference-Format}
\bibliography{main}










\end{document}